\documentstyle[epsf]{l-aa}
\begin {document}

\thesaurus{03.13.2, 08.06.3, 08.09.2 HD 2724, 08.15.1, 08.22.2.} 

\title{Simultaneous intensive photometry and high resolution spectroscopy of
$\delta$ Scuti stars}

\subtitle{III. Mode identifications and physical calibrations in HD 2724
\thanks{This work is based on observations performed at the European Southern
Observatory (ESO), La Silla, Chile.}}

\author{M.~Bossi\inst{1}, L.~Mantegazza\inst{1} and N.S.~Nu\~{n}ez\inst{1,2}}
\offprints{M.~Bossi. Internet: bossi@merate.mi.astro.it}
\institute{Osservatorio Astronomico di Brera, Via Bianchi 46, I-23807 Merate LC,
Italy.
\and
Observatorio Astron\'{o}mico P. Buenaventura Suarez, C.C. 1129, Asunci\'{o}n,
Paraguay.}

\date{ Received date; accepted date}

\maketitle

\markboth{Bossi {\em et al.}: Mode determinations and physical calibrations in
HD 2724.}{}

\begin{abstract}

On the basis of our new simultaneous photometry and spectroscopy (885 $uvby$
differential measurements in 11 nights and 154 spectrograms of the
FeII$\lambda$4508 region in 5 nights), we can detect 12 probable
periodicities in the variability pattern of this star, determining the
frequencies of 7 without any ambiguity. Through a direct fit of pulsational
models to our data, we estimate the inclination of rotational axis to be about
50$^{\circ}$ and get a reliable identification of 4 modes as well as
useful bits of information about the others: no retrograde mode is visible,
whereas the star seems to show a certain preference for purely sectorial
prograde oscillations. Finally, the attribution of our lowest frequency to the
radial fundamental pulsation allows a new calibration of physical parameters. In
particular, the gravity can be determined with unusual accuracy and the
luminosity evaluation becomes more consistent with the Hipparcos astrometry.


\end{abstract}

\section{Introduction}

The serendipitous discovery of HD 2724 as a variable star is due to Reipurth
(1981), which chose it as one of the comparison objects for his differential
photometry of the eclipsing binary AG Phe. The author identified HD 2724 as a
probable $\delta$ Scuti star and guessed a tentative period of
0$^{d}$.174 ({\em i.e.} a frequency of 5.75 d$^{-1}$). Lampens (1992)
met this periodicity again analysing her excellent sequences of absolute
measurement obtained at La Silla in 1984-85 by means of the
UBVB$_{1}$B$_{2}$V$_{1}$G Geneva photometer. Lampens' analysis
shows multiperiodic variations which are typical of the $\delta$ Scuti
light curves: besides the above mentioned frequency, she identified
unambiguously another component at $\sim$7.38 d$^{-1}$ and suggested
$\sim$6.50 d$^{-1}$ and $\sim$4.34 d$^{-1}$ as two additional
candidate frequencies.

HD 2724 is classified as an F2 III star in Hoffleit and Jaschek (1982). Lampens
(1992) gets from her photometry T$_{eff}$ = 7180$^{\circ}$K and
M$_{V}$ = 0.93. Physical parameters can be evaluated also by using the
$uvby\beta$ colours published by Hauck \& Mermilliod (1990). Moon's \&
Dworetsky's (1985) grids lead us to estimate T$_{eff}$ and $\log g$ at
7280$^{\circ}$K and 3.56 respectively, while Villa \& Breger (1998) obtain
from their still unpublished calibration, based on Canuto's \& Mazzitelli's
(1991) models and performed using dereddened indices, T$_{eff}$ =
7216$^{\circ}$K and $\log g$ = 3.64. As to the absolute magnitude,
Crawford's (1979) calibration yields M$_{V}$ = 1.12, E(b-y) = 0.014 and
therefore A$_{V}$ = 0.060. Nevertheless, our photometric evaluations of
luminosity are now to be revised owing to new astrometric data: in the Hipparcos
Satellite General Catalogue (ESA, 1997), this object (HIC 2388) appears with a
parallax $\pi$ = 7.77 $\pm$.72 mas, which, taking account of the above
assessed interstellar extinction, corresponds to an absolute magnitude
M$_{V}$ = 0.57 $\pm$.20. In principle, pulsational masses could help us
to adjust these calibrations. It would entail, however, a thorough knowledge of
pulsational states, which today might be achieved only by combining photometry
and spectroscopy in a synergetic approach (see {\em e.g.} Bossi {\em et al.},
1994, or Mantegazza {\em et al.}, 1998).

In order to exploit the complementarity between photometry and spectroscopy in
studying dynamical processes like stellar pulsations, we are performing for many
years simultaneous observational campaigns of $\delta$ Scuti stars through
both these techniques (Mantegazza {\em et al.}, 1994; Mantegazza \& Poretti,
1996). The present work on HD 2724 falls within this frame.

\section{Observations and data processing}

Both our photometry and spectroscopy have been performed at the ESO Observatory
(La Silla, Chile) in September 1993.

\begin{figure*}
\epsfysize=17.5truecm
\epsffile{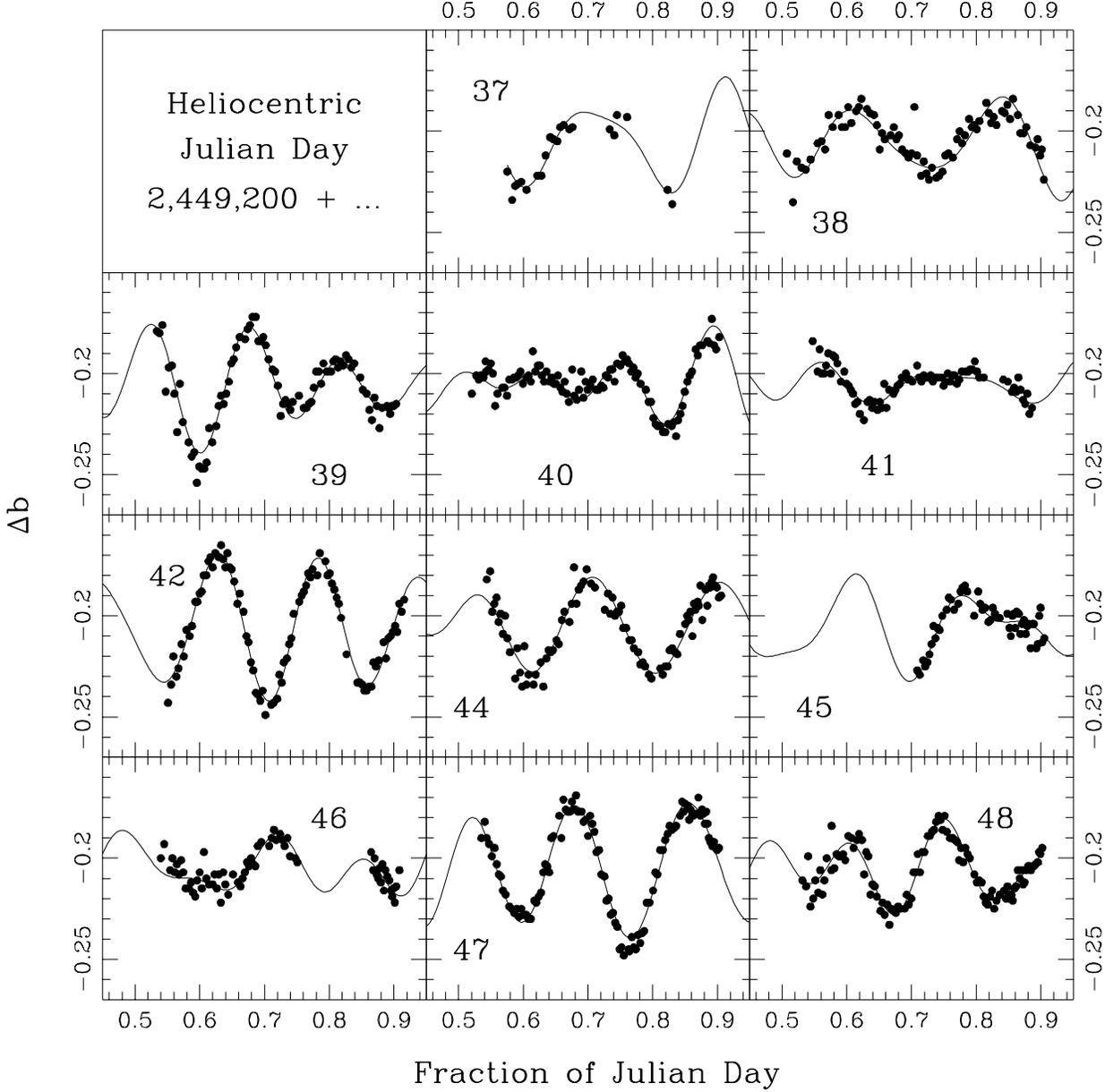}
\caption{$\Delta$b light curve of HD 2724 (dots) compared with its best
fitting 10 sinusoid model (solid line) as shown in Tab.2; the luminosity grows
upward.}
\end{figure*}

Our photometric data consist of 885 $uvby$ differential measurements
obtained through the Danish 0.5 m telescope and spread over 11 almost
consecutive nights with a baseline of $\sim$11 days, the total useful
observation time amounted to $\sim$90 hours. As comparison stars we employed
HD 3136, HD 1683 and HD 1856. All the magnitude and colour differences are given
with respect to the reference star HD 3136: our large set of data allowed us to
do it by shifting all the comparison objects to the same ficticious magnitudes
with great accuracy. Considering the size of our observational field (HD 1683
is about 4$^{\circ}$ to the NW of HD 3136), the data have been processed
using the technique developed by Poretti and Zerbi (1993) in order to deal with
the variations in extinction coefficients during each night.

\begin{table}
\caption{Statistical parameters characterizing our differential light curves
of HD 2724 and their reliability.}
\begin{tabular}{lcccc}
\hline
 & & & & \\
 & $\Delta$u & $\Delta$v & $\Delta$b & $\Delta$y \\
 & & & & \\
\hline
 & & & & \\
~Standard deviation (mmag) & 16.8 & 18.9 & 15.9 & 13.6 \\
~White noise (mmag) & 9.2 & 5.9 & 4.7 & 4.9 \\
~S/N ratio (dB) & 4 & 10 & 10 & 8 \\
~Ck1 Stand. Dev. (mmag) & 8.9 & 5.4 & 4.9 & 5.4 \\
~Ck2 Stand. Dev. (mmag) & 8.0 & 5.3 & 5.0 & 5.0 \\
 & & & & \\
\hline
\end{tabular}
\end{table}

The basic statistical parameters which characterize the resulting light curves
of HD 2724 and their reliability are presented in Tab.1, where the white noise
content of each time series has been evaluated from the root--mean--square
difference between closely consecutive data. In every light--band, the standard
deviations of the HD 1683 -- HD 3136 (Ck1) and HD 1856 -- HD 3136 (Ck2) curves,
indistinguishable from the corresponding white noises, assure the stability of
our comparison stars.

As expected in a $\delta$ Scuti star, we got the best signal--to--noise
ratios in the $\Delta$v and $\Delta$b series. The $\Delta$b light
curve is displayed in Fig.1 (the dots correspond to individual measurements).

Simultaneously with our photometric measurements, we performed spectroscopic
observations using the Coud\'{e} Echelle Spectrograph (CES) attached to the
Coud\'{e} Auxiliary Telescope (CAT) and equipped with a CCD detector. The
adopted configuration and the exposure times, which varied between 10 and 15
minutes, gave a wavelength resolution of $\sim$0.075 \AA~with a sampling of
$\sim$0.036 \AA~in the range 4490--4526 \AA~with an average
signal--to--noise ratio at the continuum level of $\sim$47 dB ({\em i.e.},
in a linear scale, a ratio of 1 to $\sim$233 between the noise standard
deviation and the continuum flux). The resulting 154 spectrograms, distributed
over 5 consecutive nights with a total useful observing time of about 34 hours,
have been processed using the MIDAS package developed by the ESO.

\begin{figure}
\epsfysize=8.5truecm
\epsffile{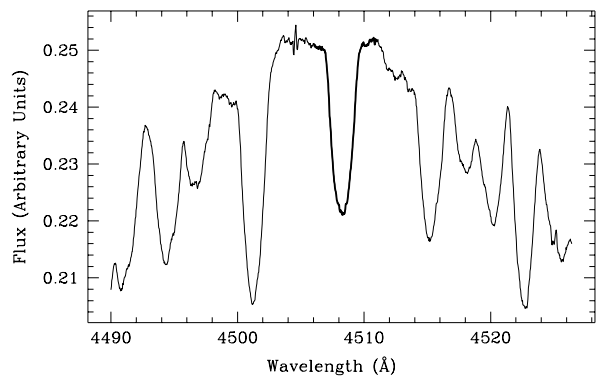}
\vskip -2.8cm
\caption{The average of our 154 spectrograms; the FeII line at $\sim$4508
\AA~ has been prominently displayed.}
\end{figure}

The mean spectrum displayed in Fig.2 shows that the star is a moderately fast
rotator. In the observed region, the blending of adjacent features spares the
sole FeII line at 4508.29 \AA: only this line borders on stretches of spectral
continuum enough to allow its stable normalization.

The average FeII$\lambda$4508 line profile resulting from the normalization
to the continuum flux is compared in the top of Fig.3 with its best fitting
synthetic profile obtained convolving together the instrumental profile, a
gaussian intrinsic profile corresponding to an atomic iron gas at
7,200$^{\circ}$K and a rotational profile. This procedure yields an estimate
of the Doppler broadening v$\sin$i = 82 km s$^{-1}$ with an uncertainty
of the order of 2 km s$^{-1}$.

\begin{figure}
\epsfysize=8.5truecm
\epsffile{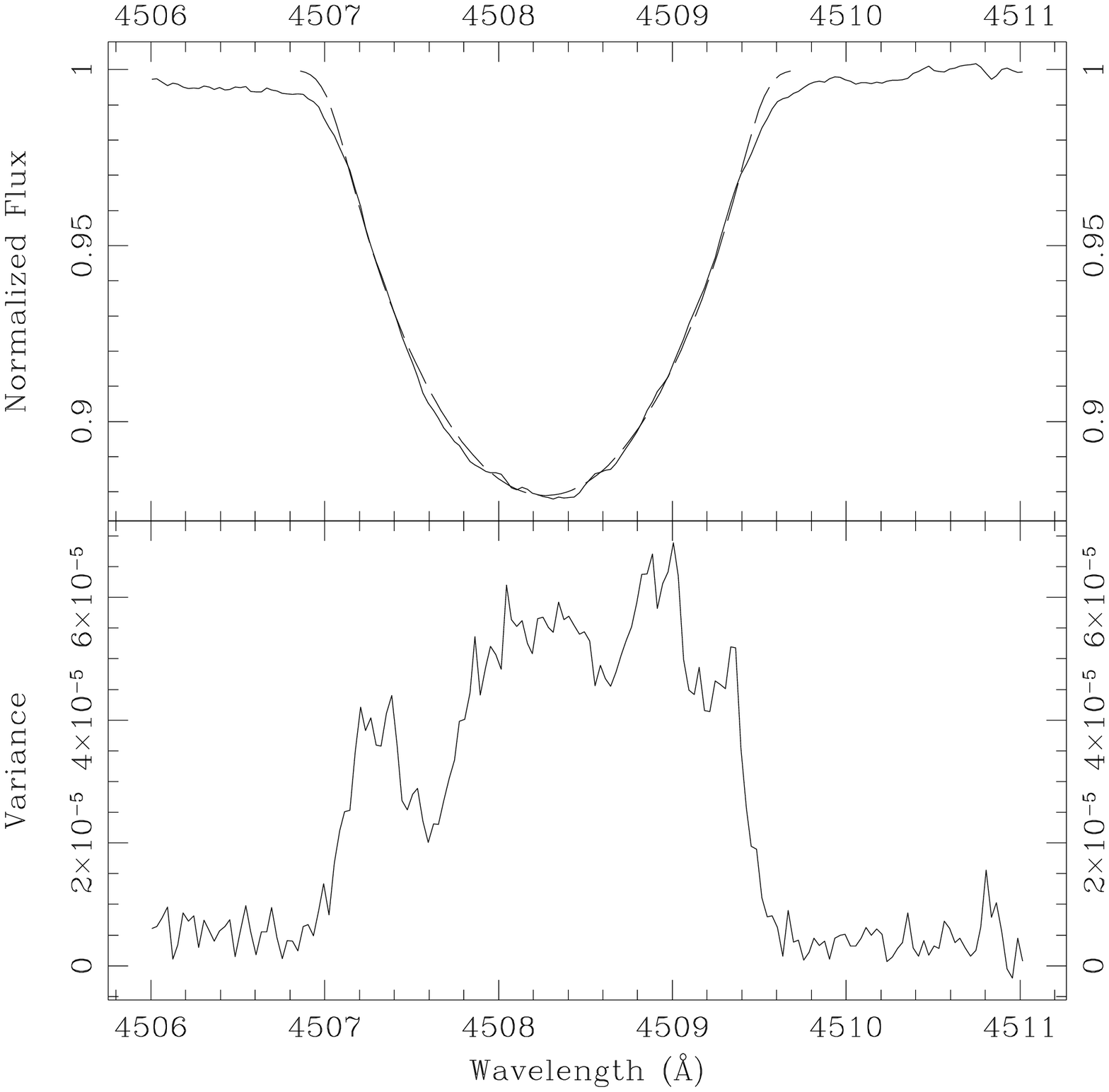}
\caption{The FeII line region at $\sim$4508 \AA: average of our 154
spectrograms normalized to the continuum flux (top, solid line); the best
fitting synthetic profile corresponding to v$\sin$i = 82 km s$^{-1}$
(top, dashed line); variance of the signal as a function of the wavelength
(bottom).}
\end{figure}

In the bottom of the same Fig.3 we present, as a function of the wavelength, the
variance of the sequences consisting of the successive flux values (normalized
to the stellar continuum) registered at each pixel. These variances have been
previously purified of the white noise contributions, which have been evaluated
pixel by pixel determining, like we did in the analysis of our photometric
series, the root--mean--square differences between closely consecutive data.
Apparently, the star shows a considerable spectral variability: the height of
the main variance peak indicates variations of flux whose standard deviation
exceeds the 0.8\% of the continuum level.

\section{Frequency analysis}

The frequency analysis of our photometric sequences has been performed both by
using the CLEAN algorithm (Roberts {\em et al.}, 1987) and through the
least--square sine--fitting technique (Vanicek, 1971; Mantegazza
{\em et al.}, 1995). At each step of Vanicek's procedure, the outcomes have been
systematically adjusted by means of simultaneous nonlinear least squares fits.
These methods made us able to detect the presence of at least 10 periodic
components in our light curves. An outline of the best fitting 10 frequency
model for the $\Delta$b curve is presented in Tab.2. The fitting curve
(solid line) can be visually compared with the measurements (dots) in Fig.1. 
Also a comparison between the root--mean--square residual (4.9 mmag) and the
corresponding white noise (4.7 mmag) shows this solution to represent the light
variations quite reasonably. Nevertheless, we are allowed to consider only 7
frequencies as completely reliable: if we examine different curves
($\Delta$y, $\Delta$v or $\Delta$u) and/or if we model them with a
different number of periodicities (9 or 11), we can improve some fits replacing
$\nu_{2}$, $\nu_{4}$ and/or $\nu_{10}$ with values at a distance of
$\sim$1 d$^{-1}$. We have to point out that $\nu_{2}$ and
$\nu_{4}$ are close to the resolution limit respectively from $\nu_{1}$
and $\nu_{5}$, {\em i.e.} from the first and the third strongest components
of the observed variability. Obviously, this fact entangles our frequency
spectra hindering us in the identification of the significant peaks. On the
other hand, no doubt either these periodicities or the alternative ones at
$\pm$1 d$^{-1}$ are present in our signals: tests performed using
synthetic data (Bossi \& Nu\~{n}ez, in preparation) made us sure that our
procedures permit frequency resolutions substantially better than the
conventional ones. Anyway, in consequence of these ambiguities, we must regard
the errors presented in Tab.2 as lower bounds. They have been determined by
means of the standard statistical approach assuming the observed variability to
consist just of the 10 considered sinusoids and white noise: the replacement of
one of these components with an alternative periodicity results in affecting the
determination of all the other parameters. Another series of tests showed us,
for example, how a prudential evaluation of the frequency uncertainties, which
should take aliasing phenomena into account, would increase the error bars
associated with unambigous components to $\sim$0.01 d$^{-1}$.

\begin{table}
\caption{The best fitting 10 sinusoid model of the $\Delta$b light curve;
the phases are referred to t$_{0}$ = H.J.D. 2,449,243.6950.}
\begin{tabular}{rccl}
\hline
 & & & \\
~~~Comp. & Frequency & Amplitude & ~~Phase \\
 & (d$^{-1}$) & (mmag) & ~~(rad) \\
 & & & \\
\hline
 & & & \\
$\nu_{1}$~~ & 4.430 $\pm$.002 & 8.4 $\pm$.3 &
-1.14 $\pm$.04~~~~~ \\
$\nu_{2}$~~ & 4.536 $\pm$.008 & 2.8 $\pm$.3 & ~3.06 $\pm$.11 \\
$\nu_{3}$~~ & 5.311 $\pm$.002 & 6.7 $\pm$.3 & ~0.21 $\pm$.04 \\
$\nu_{4}$~~ & 5.629 $\pm$.007 & 2.9 $\pm$.3 & -3.04 $\pm$.11 \\
$\nu_{5}$~~ & 5.736 $\pm$.001 & 11.1 $\pm$.3 & ~2.28 $\pm$.03 \\
$\nu_{6}$~~ & 5.877 $\pm$.003 & 3.9 $\pm$.3 & -2.25 $\pm$.08 \\
$\nu_{7}$~~ & 6.123 $\pm$.002 & 5.6 $\pm$.3 & -0.78 $\pm$.05 \\
$\nu_{8}$~~ & 6.488 $\pm$.003 & 4.8 $\pm$.3 & ~1.87 $\pm$.06 \\
$\nu_{9}$~~ & 7.382 $\pm$.001 & 9.1 $\pm$.3 & -0.34 $\pm$.03 \\
$\nu_{10}$~~ & 8.049 $\pm$.003 & 3.6 $\pm$.3 & -1.64 $\pm$.07 \\
 & & & \\
\hline
\end{tabular}
\end{table}

Reipurth's (1981) periodicity is recognizable as $\nu_{5}$ and also the
frequencies indicated by Lampens (1992) with the surprising number of 5 decimal
places lie close to $\nu_{1}$, $\nu_{5}$, $\nu_{8}$ and
$\nu_{9}$. In order to check if the pulsational status of this star is
stable in time scales of years, Lampens' measurements might deserve a further
analysis. In fact, those data permitted the detection of four substantially
correct frequencies in spite of their unfortunate distribution and in spite of
the rough prewhitening technique adopted by the author: if we resort to this
procedure in the presence of complex multiperiodic behaviours, sensible fit
errors accumulate step by step in the successively analysed residuals,
eventually leading us on the wrong track.

The analysis of photometric data is not bound to exhaust the search for
detectable pulsations: the presence of high $\ell$--degree modes affects
the spectral line profiles, especially if broadened by an high projected
rotational velocity, more than the observed luminosity. Therefore, a period
analysis of the FeII$\lambda$4508 profile variations has been performed
pixel by pixel using again both Vanicek's (1971) method (see also Mantegazza,
1997) and the CLEAN algorithm. The top panel of Fig.4 shows the resulting CLEAN
spectrum as a function of the wavelength; moving to the bottom, we meet the
corresponding mean spectrum and finally we can compare it with a photometric
power spectrum drawn according to the model of Tab.2. Vanicek's outcomes reveal
the presence of few more photometric frequencies in our spectroscopic series,
but they are not very practical for graphic presentation: each detected
component would need its figure.

\begin{figure}
\epsfysize=8.5truecm
\epsffile{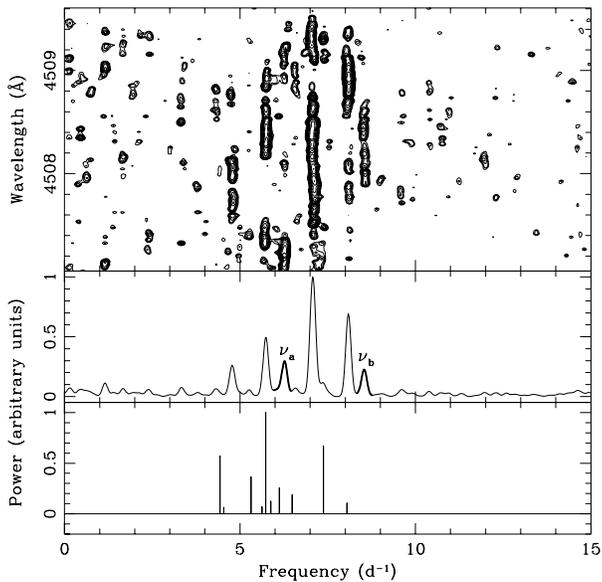}
\caption{From the top to the bottom: the CLEAN spectrum of the flux series
describing pixel by pixel the evolution of the FeII$\lambda$4508 line
profile shown as a function of the wavelength, the corresponding mean spectrum
and the photometric power spectrum which results from the model of Tab.2.}
\end{figure}

The inadequacy of our spectroscopic data window is apparent: a) $\nu_{5}$
and $\nu_{10}$ switch between two alternative frequency values at a distance
of 1 d$^{-1}$ changing the wavelength, {\em i.e.} the phases of the
corresponding oscillations; b) the spectral resolution would not be enough for
separating frequencies close to each other like $\nu_{4}$, $\nu_{5}$ and
$\nu_{6}$. Anyway, as it was only to be expected, pulsations manifest
themselves in photometric and spectroscopic variations with independent
intensities: $\nu_{10}$, which in Tab.2 shows the third last amplitude, is
beyond all doubt the dominant spectroscopic periodicity. Besides the above
quoted $\nu_{10}$ and $\nu_{5}$, also the photometric frequency
$\nu_{9}$ can be immediately recognized in the bump on the right of the main
peak of our CLEAN mean spectrum. The components at $\sim$6.27
($\nu_{a}$) and $\sim$8.55 d$^{-1}$ ($\nu_{b}$), prominently
displayed in the same panel of Fig.4 and invisible in the light curves, are
likely to correspond to real periodicities too: they do not disappear even
assuming $\nu_{7}$ and $\nu_{8}$ as known constituents in Vanicek's
analysis. Finally, Vanicek's (1971) approach allows us, adding $\nu_{3}$
and $\nu_{8}$, to complete our set of spectroscopic frequencies.

\section{Mode identification}

The technique developed by Garrido {\em et al.} (1990) in order to discriminate
between radial and low $\ell$--degree non--radial pulsations in $\delta$
Scuti stars following Dziembowski (1977) and Watson (1988) could help us, in
principle, with the characterization of our 10 photometric modes. Nevertheles,
phase uncertainties comparable with the expected lags hinder us in quarrying in
our Str\"{o}mgren curves for useful bits of information. Fig.5 displays
{\em e.g.} the phase shifts between the v and the y variations shown by each of
these components: we cannot get from this kind of diagram much more than a
marginal evidence supporting the radial character of $\nu_{1}$ suggested
also by its merely photometric relevance. On the other hand, the absence of
$\nu_{a}$ and $\nu_{b}$ in our photometric signal implies relatively
high $\ell$--values for them.

\begin{figure}
\epsfysize=8.5truecm
\epsffile{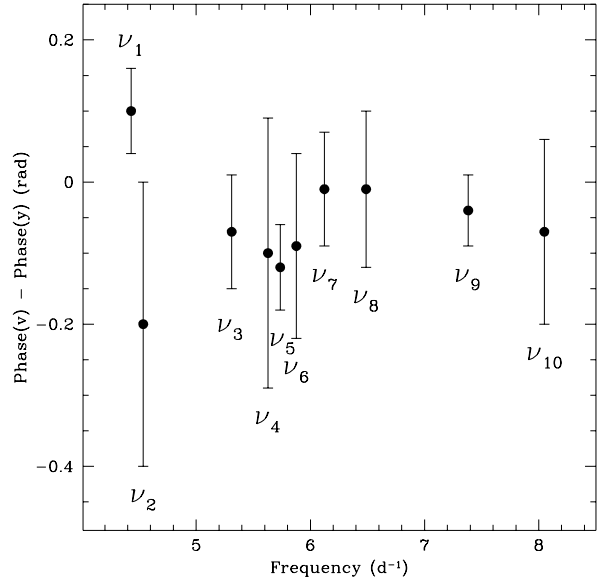}
\caption{Phase shifts between v and y shown by the periodic components of our
light curves.}
\end{figure}

At this point, we could not go any farther without resorting to some line
profile analysis. The most common tools to perform such computer--limited task
in reasonable calculating times are currently the moment method proposed by
Balona (1987) and the two--dimensional Fourier analysis combined with the
Doppler imaging by Kennelly {\em et al.} (1992). The last approach is in
principle the most suitable one for the study of rapidly rotating objects like
HD 2724: a dominant Doppler broadening makes the spectral lines useful as
one--dimensional maps of the stellar surfaces. This ought to permit at least
the immediate determination of the $m$--numbers. Nevertheless, as shown by
Telting \& Schrijvers (1995), to rely mechanically on this technique is
dangerous: under certain conditions ($\mid m \mid~<~\ell;~i~\ll~90^{\circ}\)),
its unthinking use leads us to overestimate $\mid m \mid$. On the other
hand, Balona's (1987) procedure is based on a series of strong simplifications
perhaps today less and less necessary due to the continuous advance of computer
technology.

For those reasons, at Balona's (1997) suggestion, we nerved ourselves to perform
a direct fit of pulsational models to our line profile variations taking also
the photometric measurements into account.

\begin{enumerate}

\item The first step was a simultaneous least--square fit performed pixel by
pixel adjusting the sum of a constant parameter plus 7 sinusoids to each
sequence of normalized fluxes. We assumed the frequencies of the periodic terms
to be the ones singled out in the preceding section as relevant to the
spectroscopic variability: $\nu_{3}$, $\nu_{5}$, $\nu_{8}$,
$\nu_{9}$, $\nu_{10}$, $\nu_{a}$ and $\nu_{b}$. The set of
constant parameters describes the unperturbed line profile, while each group of
sinusoids draws the corresponding periodic component of the profile variations.
The pattern of the main photometric component $\nu_{5}$ ($\sim$5.74
d$^{-1}$), quarried out of our data by means of this procedure, is shown for
example in Fig.6 (solid lines).

\begin{figure}
\epsfysize=8.5truecm
\epsffile{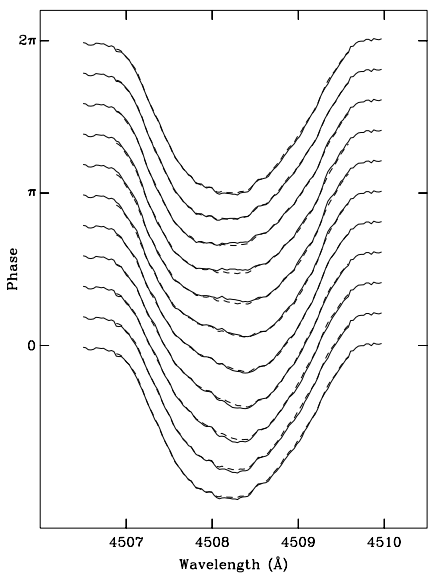}
\caption{The component of the line profile variability corresponding to the
frequency $\nu_{5}$ $\simeq$ 5.74 d$^{-1}$ (solid lines), compared
with the best fitting synthetic model (dashed lines), corresponding to
$\ell$=2, $m$=-2, $i$=50$^{\circ}$.}
\end{figure}

\item We assumed as fixed input parameters a r.m.s. intrinsic line width 
W$_{i}$ = 11.4 Km s$^{-1}$, a Doppler broadening v$\sin$i = 81.7 Km
s $^{-1}$ and a limb darkening coefficient $u_{\lambda 4509}$ =
0.7: the values of W$_{i}$ and of the projected rotational velocity resulted
from a simultaneous non--linear least--square fit performed on the unperturbed
profile assuming a gaussian intrinsic line shape; for the limb darkening
see {\em e.g.} D\'{\i}az--Cordov\'{e}s {\em et al.} (1995). As regards the
inclination $i$ of the rotational axis, we scanned the range
30$^{\circ}$--90$^{\circ}$ with a resolution of 7.5$^{\circ}$
($i \le 22.5^{\circ}$ would entail an equatorial velocity of at least 215
Km s$^{-1}$).

\item Each of the 7 above quoted variability components has been described
putting 10 instantaneous line profiles, with phase intervals of $\pi$/5 like
in Fig.6, together with the corresponding photometric pattern defined in Tab.2.
Obviously, the photometric amplitude assigned to the purely spectroscopic
components $\nu_{a}$ and $\nu_{b}$ was zero. Then, 10 synthetic profiles
and a synthetic light curve have been adjusted to these data by means of a
non--linear least--square technique for each possible candidate mode, {\em i.e.}
for each plausible couple of ($\ell;~m$) numbers, and for each explored 
value of the $i$ parameter. Both the spectroscopic and photometric models
have been produced through the LNPROF code, kindly put at our disposal by L.A.
Balona. Balona's algorithm uses a first order expansion in the ratio
$\Omega/\omega$ of the rotational frequency to the pulsational one: a
comparison of $\Omega$ $\simeq$ 0.5/$\sin i$ d$^{-1}$ to our
7 scrutinized frequencies makes this approximation quite reasonable. Finally,
assuming the presence of $p$ modes and operating in adiabatic
approximation, we could link the horizontal displacement velocity $V_{h}$
with the vertical one $V_{r}$ through the relation
$V_{h}~\simeq~74.4~Q^{2}V_{r}$. Thus, only 4 free parameters remained to be
determined for each fixed $\ell$, $m$ and $i$: the amplitudes and
phases of the luminance modulation on the stellar surface and of $V_{r}$.
These parameters have been adjusted combining the mean--square differences
between observational and synthetic models in one discriminant which assigns
equal weights to photometry and spectroscopy, and minimizing it through the
iterative downhill simplex method (Press {\em et al.}, 1992). In Fig.6, for
example, we can compare the best fitting synthetic profiles which describe the
component $\nu_{5}$ of the spectral variability (dashed lines) to the
observational ones (solid lines). The resulting discriminant values are shown in
Fig.7 for each frequency and for the best fitting candidate couples
($\ell; m$) as functions of the inclination angle.

\begin{figure*}
\epsfysize=17.5truecm
\epsffile{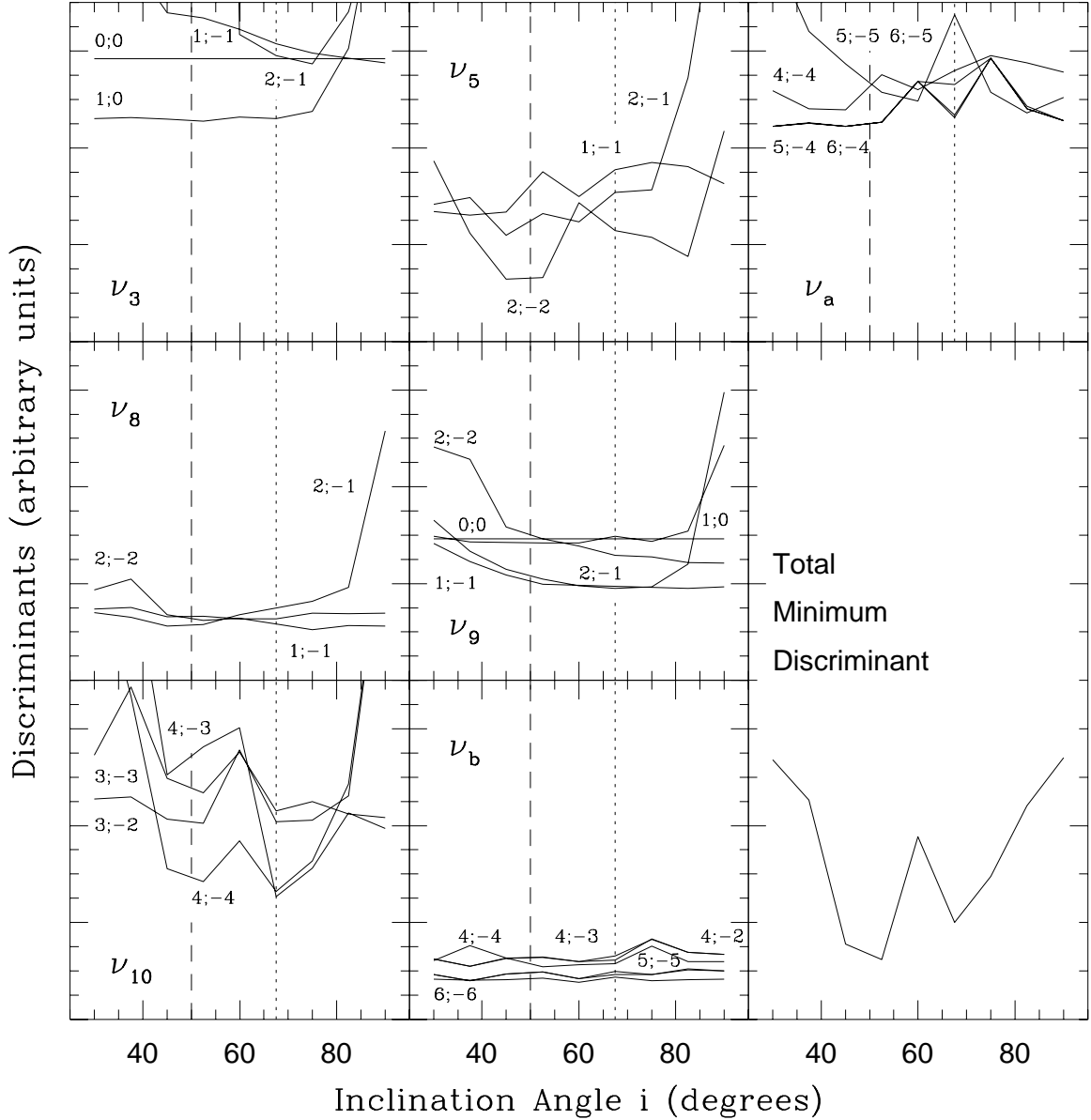}
\caption{Minimum discriminant values, correponding to the best fitting candidate
couples ($\ell;m$), shown as functions of the inclination angle for each
frequency which we could detect in the spectral variability. The panel visible
low on the right shows, in the same scale and again as a function of $i$,
the sum of the 7 minimum discriminants.}
\end{figure*}

\item In order to get information about the plausible inclination of the
rotational axis, taking our bearing among the pulsational modes, we computed,
again as a function of $i$, the sum of the minimum discriminants
corresponding to the 7 frequencies which we could detect in the spectral
variability (see the panel visible low on the right in the same figure).

\end{enumerate}

\begin{table*}
\begin{center}
\caption{Mode identification obtained combining photometric and spectroscopic
information and accepting the most probable inclination
$i~\simeq~50^{\circ}$. The third column shows the corresponding frequencies
in a co--rotating reference frame obtained assuming a rotational frequency
$\Omega$ = 0.45 $\pm$.03 d$^{-1}$.}
\begin{tabular}{cll}
\hline
 &  & \\
~~~~~~~~~~~Component~~~~~~~~~~~~~ & Plausible $\ell;m$ &
~~~~~~~~~~~Proper frequency (d$^{-1}$) \\
 &  & \\
\hline
 &  & \\
$\nu_{1}$ & 0;0 & ~~~~~~~~~~~~~4.43 $\pm$.01 \\
$\nu_{2}$ & non--radial mode & \\
$\nu_{3}$ & 1;0 & ~~~~~~~~~~~~~5.31 $\pm$.01 \\
$\nu_{4}$ & non--radial mode & \\
$\nu_{5}$ & 2;-2 & ~~~~~~~~~~~~~4.84 $\pm$.07 \\
$\nu_{6}$ & 0;0? & ~~~~~~~~~~~~~5.88 $\pm$.01? \\
$\nu_{7}$ & non--radial mode? & \\
$\nu_{a}$ & 5;-4 ~~or~~ 6;-4? & ~~~~~~~~~~~~~4.48 $\pm$.13 \\
$\nu_{8}$ & 2;-1, 2;-2 ~~or~~ 1;-1 & ~~~~~~~~~~~~~6.04 $\pm$.03 ~~or~~
5.60 $\pm$.07 \\
$\nu_{9}$ & 1;-1 ~~or~~ 2;-1 & ~~~~~~~~~~~~~6.94 $\pm$.03 \\
$\nu_{10}$ & 4;-4 & ~~~~~~~~~~~~~6.27 $\pm$.13 \\
$\nu_{b}$ & 6;-6, 5;-5 ~~or~~ 4;-4 & ~~~~~~~~~~~~~5.88 $\pm$.20 ~~or~~
6.32 $\pm$.16 ~~or~~ 6.77 $\pm$.13~~~~~~~~~~~~~ \\
 & & \\
\hline
\end{tabular}
\end{center}
\end{table*}

The inclination angle $i$ results likely to be close to 50$^{\circ}$,
even if an alternative value between 65$^{\circ}$ and 70$^{\circ}$
cannot be ruled out. In Fig.7, these angles are indicated by the vertical dashed
line and by the dotted one respectivlely: it is easy to notice how the choice of
the first value, besides minimizing the r.m.s. residual, to some extent
disentangles the mode identification. As regards the pulsational modes, our
information, considering also the photometric phase shifts, is summarized in
Tab.3. Despite the incompleteness of these results, an item deserves a certain
interest: we detected no retrograde perturbation. No way it can be a simple
observational bias due to the fast rotation of HD 2724: the sign of $m$ is
defined by our code in a co--rotating reference frame. If we might hazard a
second more uncertain guess, we would also say this class of stars prefers,
among the prograde non--radial pulsations, the purely sectorial ones.

\section{Physical parameters}

The representative points of HD 2724 in the plane ($\log g;~T_{eff}$) which
correspond to the calibrations of Moon \& Dworetsky (1985) and of Villa \&
Breger (1998) are shown in Fig.8 respectively by the black dot and by the open
circle. In the same figure, assuming Hipparcos' parallax, the solid line
indicates the gravity--temperature combinations which associate our lowest
frequency $\nu_{1}$ with a pulsational constant Q = Q$_{0}$ =
0$^{d}$.033, {\em i.e.}, according to Fitch (1981), whith the fundamental
radial pulsation. Finally, the identification of $\nu_{1}$ as a g$_{1}$
mode with $\ell$ = 2 and Q = 0$^{d}$.051 ({\em cf.} again Fitch, 1981)
corresponds in the figure to the dashed line. We have no reasonable alternative
but these two possible Q values. For one thing, in fact, the identification of
$\nu_{1}$ as an high--$\ell$ g--mode is very improbable: considering its
photometric amplitude, an high degree non--radial oscillation would be easily
detected in our spectrograms. For another thing, both radial overtones and
non--radial p--modes would entail unrealistic pulsational constants Q $<$
Q$_{0}$.

\begin{figure}
\epsfysize=8.5truecm
\epsffile{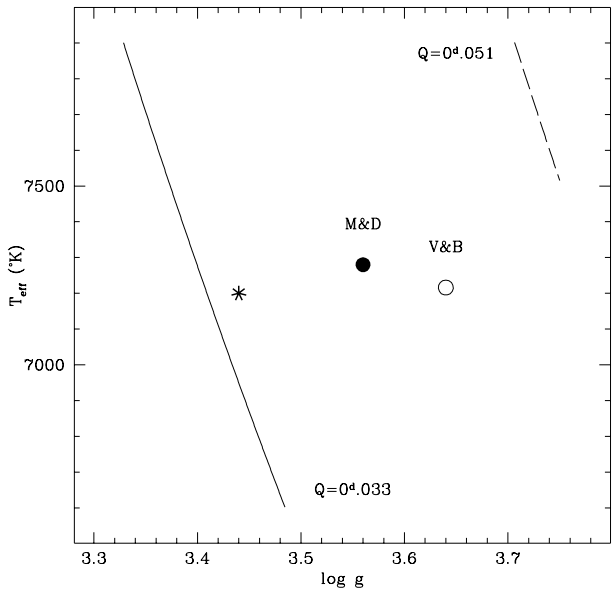}
\caption{Representative points of HD 2724 in the plane ($\log g;~T_{eff}$)
according to the calibration of Moon \& Dworetsky (black spot), to Villa's \&
Breger's (1998) method (open circle) and to the results shown in the present
work (asterisk). Assuming Hipparco's parallax, the solid line corresponds to the
identification of $\nu_{1}$ as the fundamental pulsation, the dashed one as
a g$_{1}$ mode with $\ell$ = 2.}
\end{figure}

Combined with the evolutionary models of Shaller {\em et al.} (1992), the
association of $\nu_{1}$ with Q = 0$^{d}$.051 does not allow us to
reconcile photometry with astrometry. This choice, characterizing our object as
a 2 M$_{\odot}$ star with T$_{eff}$ = 7200$^{\circ}$K, $\log g$
= 3.67 and M$_{V}$ = 1.1, would lead us back to Crawford's (1979) magnitude
calibration, at a $\sim 3 \sigma$ distance from Hipparco's one.

\begin{figure}
\epsfysize=8.5truecm
\epsffile{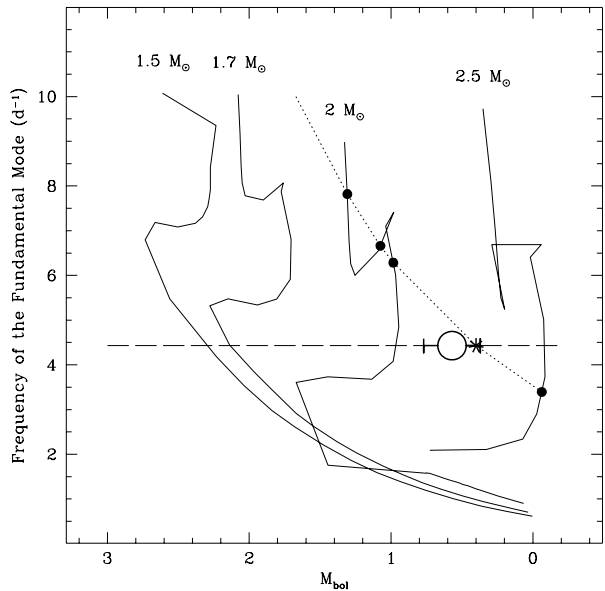}
\caption{Evolutionary tracks for different mass values translated into the plane
Bolometric Magnitude -- Fundamental Frequency (solid lines), the dots indicate
the instants of T$_{eff}$ = 7200$^{\circ}$K, the horizontal dashed line
corresponds to our frequency $\nu_{1}$. The representative points of HD 2724
according to the Hipparcos parallax and to our calibration are shown
respectively by the open circle with horizontal error bar and by the asterisk.}
\end{figure}

Therefore, we can add the astrometric evidence to our previous indications
({\em cf.} the last section) in support of the radial hypothesis, we can
identify $\nu_{1}$ as the fundamental mode and have a try assuming only an
effective temperature of 7200 $\pm$100$^{\circ}$K and a pulsational
constant Q$_{0}$ = 0$^{d}$.033 with a theoretical uncertainty of the
order of 0$^{d}$.001 (Fitch, 1981). The evolutionary tracks of Schaller
{\em et al.} (1992) can be easily translated, like in Fig.9, from the plane 
($\log$(L/L$_{\odot}$); $\log$T$_{eff}$) into the plane
(M$_{bol}$; $\nu_{0}$), where $\nu_{0}$ is the frequency of the
fundamental mode determined assuming Q$_{0}$ = 0$^{d}$.033. The instants
that the stars, following downwards their evolutionary paths (solid lines),
reach an effective temperature T$_{eff}$ = 7200$^{\circ}$K are indicated
by dots: for M = 1.7 M$_{\odot}$ it occurs outside the panel, while a star
with M = 1.5 M$_{\odot}$ keeps always below this temperature. Finally, an
horizontal dashed line corresponds to our frequency $\nu_{1}$. This picture
allows us to assign to HD 2724, which we are apparently observing during its
expansion immediately after the exhaustion of the hydrogen burning, a mass of
about 2.25 M$_{\odot}$, a bolometric absolute magnitude not far from 0.4 and
a surface gravity $\log g$ $\simeq$ 3.44. The asterisks in Fig.8 and
Fig.9 show the corresponding representative points respectively in the plane
($log g;~ T_{eff}$) and (M$_{bol}$; $\nu_{0}$). In Fig.9, according
to Hipparcos parallax, we ought to locate it in the position indicated by the
open circle with horizontal error bar: since the absolute magnitude value was
not among our assumptions, neither the closeness of these points nor, in Fig.8,
the adjacency of the star to the line corresponding to Q$_{0}$ were to be
taken for granted {\em a priori}.

\begin{table}
\caption{Basic physical parameters of HD 2724 according to our pulsational
model.}
\begin{tabular}{ll}
\hline
 & \\
~~~Parameter & Estimated value~~~~~ \\
 & \\
\hline
 & \\
~~~Mass & 2.25 $\pm$.10 M$_{\odot}$ \\
~~~Effective temperature & 7200 $\pm$100$^{\circ}$K \\
~~~Absolute bolometric magnitude & 0.4 $^{+.2}_{-.1}$ \\
~~~$\log g$ & 3.44 $\pm$.03 \\
~~~Radius & 4.7 $\pm$.3 R$_{\odot}$ \\
~~~Rotational frequency & 0.46 $\pm$.03 d$^{-1}$ \\
 & \\
\hline
\end{tabular}
\end{table}

The consistency of these results leads us to summarize the corresponding
physical picture in Tab.4. The asymmetrical uncertainty assigned to the absolute
bolometric magnitude takes both our model and the above quoted astrometric
constraint into account.

The rotational frequency appears even lower than previously expected
({\em cf.} Sect.4, point 3), making our linear expansion in $\Omega/\omega$
wholly legitimate {\em a posteriori}. The value of $\Omega$ shown in Tab.4
has been used for evaluating the proper pulsational frequencies referred to a
co--rotating frame presented in the third column of Tab.3. If correctly
identified, the two components detected only in the line profile variations
deserve a few interesting considerations. The proper Q--value corresponding to
$\nu_{a}$ (0$^{d}$.033 $\pm$0$^{d}$.001) neither is consistent
with a p--mode nor with a g--mode: it would entail the presence in this
$\delta$ Scuti star of an high $\ell$--order f--oscillation. As to
$\nu_{b}$, its most probable de--rotated value seems to show an 1:1
resonance with the observed frequency $\nu_{6}$ which, therefore, would not
need any rotational correction: according to its Q--value of 0$^{d}$.025,
we could identify $\nu_{6}$ as the first radial overtone.

Finally, we have to notice that the pulsational calibration
\[ \log g~=~2\log Q\nu~-~0.2M_{bol}~-~2\log T_{eff}~+~12.908~~~~~, \]
\noindent if our hypothesis is correct, permits a particularly accurate gravity
determination.

\section{Open problems}

In this work we could get a few interesting results about the pulsational
pattern presented by a $\delta$ Scuti star and about its relevance to the
study of stellar structure and evolution; nevertheless, we are far from drawing
an exhausive picture of the pulsational behaviour of HD 2724 as well as of other
$\delta$ Scuti objects.

Our light curves allowed us to detect the presence of at least 10 different
periodicities, determining unambiguously no more than 7 frequencies. Apparently,
observational bases like our 11 high quality photometric nights at La Silla are
inadequate if faced by the complex variability of these stars. The last
instructive example of such complexity is provided by Breger {\em et al.}
(1998), who identify 24 frequencies in the light curve of FG Vir, discovering,
moreover, 8 additional promising candidate pulsational components and finding
evidence of considerable amplitude variations (maybe due to beats between
unresolved frequencies) affecting one of the detected periodicities in a time
scale of one year. In the above mentioned paper, Breger and his collaborators
show also the way to obtain this kind of results, which are basically the fruits
of the largest photometric multi--site campaign devoted to date to a
$\delta$ Scuti star.

The inadequacy of our data becomes dramatic on the spectroscopic side. In fact,
although the scrutiny of our spectroscopic series led us to discover two
additional high $\ell$--order non--radial pulsation invisible in the light
curves, the poor spectral window of these data did not even allow a reliable
determination of the corresponding frequencies. Besides, blending close
periodicities who could be photometrically resolved, it affected also the
subsequent mode identification. Unfortunately, it is more difficult, even if
not less important, to obtain satisfactory series of spectroscopic observations
than photometric ones: to the delicate organizational problems of a multi--site
campaign, we would have to add the hard work of elbowing our way to adequate
telescopes.

We realize that an exhaustive pulsational picture of the $\delta$  Scuti
stars might become, under these conditions, nothing more than a tantalizing
dream. Nevertheless, let us end on an optimistic note: as shown, for example,
just in this work, interesting results can be obtained also from incomplete
patterns; besides, unexpected theoretical  developments might help to simplify
our task. Chandrasekhar \& Ferrari (1991, 1992), for example, began to explore
the exciting possibility of getting selection rules of pulsational modes through
the tranfer of methods from the quantum field theory to the
general--relativistic treatement of non--radial oscillations.

\paragraph{Acknowledgements.} We are grateful to L.A. Balona and E. Poretti for
their useful suggestions, thanks are due to Dr. Balona also for allowing us to
use his LNPROF code.

\end {document}